Figures are included as postscript files at the end. These should be extracted
and stored in files 1.ps, 2.ps, 3.ps and 4.ps. The main text includes these
files with the  command. This works on a Macintosh,
but for other machines maybe better to delete all occurances of  and
print the postscript files separately.

\documentstyle [12pt]{article}
\newcommand{\nn}{\nonumber\\}

\newcommand{\be}{\begin{equation}}
\newcommand{\ee}{\end{equation}}
\newcommand{\bea}{\begin{eqnarray}}
\newcommand{\eea}{\end{eqnarray}}
\newcommand{\G}{{\cal G}}
\newcommand{\psib}{\bar{\psi}}
\newcommand{\Psib}{\bar{\Psi}}
\newcommand{\A}{\alpha}
\newcommand{\B}{\beta}
\newcommand{\C}{\gamma}
\newcommand{\k}{{\bf k}}
\newcommand{\p}{{\bf p}}
\newcommand{\q}{{\bf q}}
\newcommand{\x}{{\bf x}}
\newcommand{\y}{{\bf y}}
\newcommand{\K}{\kappa}

\newcommand{\bfp}{{\bf p}}

\newcommand{\ra}{\rangle}
\newcommand{\la}{\langle}
\newcommand{\raa}{\rangle\!\rangle}
\newcommand{\laa}{\langle\!\langle}
\newcommand{\pid}{$\pi d$}
\newcommand{\piN}{$\pi N$}
\newcommand{\piNNpiNN}{$\pi N\! N\!\rightarrow\!\pi N\!N$}

\newcommand{\NN}{$N\!N$}
\newcommand{\NNpiNN}{$N\!N\!-\pi N\!N$}
\newcommand{\BBpiBB}{$B\!B\!-\pi B\!B$}
\newcommand{\piNN}{$\pi N\!N$}
\newcommand{\pidpid}{$\pi d\!\rightarrow\!\pi d$}
\newcommand{\pppid}{$pp\rightarrow\!\pi^+ d$}
\newcommand{\NNpid}{$N\!N\!\rightarrow\!\pi d$}
\newcommand{\NNNN}{$N\!N\!\rightarrow\!N\!N$}
\newcommand{\NNND}{$N\!N\!\rightarrow\!N\Delta$}

\newcommand{\tb}[1]{\,\,\tilde{\!\!\bar{{#1}}}}

\newcommand{\bc}{\begin{center}}
\newcommand{\ec}{\end{center}}

\newcommand{\eqn}[1]{\label{#1}}
\newcommand{\eq}[1]{Eq.\ (\ref{#1})}
\newcommand{\eqs}[1]{Eqs.\ (\ref{#1})}
\newcommand{\fign}[1]{\label{#1}}
\newcommand{\fig}[1]{Fig.\ \ref{#1}}

\begin{document}
\hspace*{\fill} \begin{tabular}{ll} FIAS-R-225 \\ November, 1993 \end{tabular}
\vspace{.5cm}

\noindent
Invited talk, 14th European Conference on Few-Body Problems in Physics,
Amsterdam, The Netherlands, 23-27 August, 1993.
\vspace{-2cm}

\vspace*{5cm}

\centerline{\bf CONVOLUTION APPROACH TO THE \mbox{\boldmath $\pi N\!N$} SYSTEM}

\bigskip
\bigskip

\centerline{B.\ Blankleider}
\centerline{\em School of Physical Sciences, Flinders University of South
Australia,}
\centerline{\em Bedford Park, S.A. 5042, Australia }
\vspace{0.6cm}

\centerline{A.\ N.\ Kvinikhidze}
\centerline{\em Mathematical Institute of Georgian Academy of Sciences}
\centerline{\em   Z.Rukhadze 1, 380093 Tbilisi, Georgia}
\vspace{1.cm}

\normalsize
\centerline{\bf ABSTRACT}
\medskip

The unitary \NNpiNN\ model contains a serious theoretical flaw: unitarity is
obtained at the price of having to use an effective \piNN\ coupling constant
that is smaller than the experimental one. This is but one aspect of a more
general renormalization problem whose origin lies in the truncation of Hilbert
space used to derive the equations. Here we present a new theoretical approach
to the \piNN\ problem where unitary equations are obtained without having to
truncate Hilbert space. Indeed, the only approximation made is the neglect
of connected three-body forces. As all possible dressings of one-particle
propagators and vertices are retained in our model, we overcome the
renormalization problems inherent in previous \piNN\ theories. The key element
of our derivation is the use of convolution integrals that have enabled us to
sum all the possible disconnected time-ordered graphs. We also discuss how the
convolution method can be extended to sum all the time orderings of a connected
graph. This has enabled us to calculate the fully dressed \NN\ one pion
exchange
potential. We show how such a calculation can be used to estimate the size of
the connected three-body forces neglected in the new \piNN\ equations. Early
indications are that such forces may be negligible.

\bigskip\medskip

\centerline{\bf INTRODUCTION}
\medskip

Eighteen months ago, at the International Few-Body Conference in Adelaide,
one of us (B.B.) summarized the status of our understanding of the \NNpiNN\
system as follows \cite{FewBody_Adelaide}, "Despite the almost thirty years of
effort, we are forced to the conclusion that all current methods have serious
deficiencies and that a radical new approach may be needed before this
outstanding problem is finally resolved". In particular, it was demonstrated
that the so called "unitary \NNpiNN\ model" [2-6], which had been the state of
the art description for more than 10 years, is theoretically flawed. It is
therefore especially pleasing to be able to present, to this conference, just
such a radically new approach to the \piNN\ system, involving  convolution
integrals, which resolves the outstanding theoretical problems of the unitary
\NNpiNN\ model, and which, at this stage, promises to effectively take into
account all the diagrams of time-ordered perturbation theory.\footnote{We note
that progress has also been made in the four-dimensional relativistic sector
where recently derived covariant equations for the \piNN\ system resolve
overcounting and undercounting problems of previous formulations \cite{KB_NP}.
In this review talk, however, we shall limit the discussion to the
three-dimensional approaches of time-ordered perturbation theory.}

As it still may not be well recognized that the unitary \NNpiNN\ model is
flawed, let us first briefly repeat here the nature of the theoretical error
inherent in the model, and how this error can lead to large inaccuracies in
predictions of observables.

The origin of the problem lies in the truncation of Hilbert space used to
derive the \NNpiNN\ equations. Truncation, in this case, means that
explicit states with more than one pion are forbidden (certain
multi-pion states are, however, included implicitly through the use of
two-body,
energy-independent potentials). This truncation has serious consequences
for the renormalization of both the two-nucleon propagator and the \piNN\
vertex.
In \fig{flaw}(a) we show the \piN\ nucleon pole diagram where the
intermediate state nucleon is dressed by one-pion loops; however, the initial
and final state nucleons do not include dressing since two-pion states are
neglected in the truncation. Since close to the nucleon pole the dressed
one-nucleon propagator is of the form $g(E) \sim Z/(E-m)$, where $Z$ is
the residue at the pole, \fig{flaw}(a) illustrates how each \piNN\ vertex
$f(E)$ gets effectively renormalized by a factor of $Z^{1/2}$. Thus $f_{\pi
N\!N}
= Z^{1/2}f(m)$ is essentially the \piNN\ coupling constant, and this fact is
used
to fix the strength parameter in the form factor $f(E)$.  With all other
parameters of $f(E)$ fixed to reproduce experimental \piN\ phase shifts, this
form factor then enters the unitary \NNpiNN\ equations as an input. As
illustrated in \fig{flaw}(b), when the \NN\ one pion exchange (OPE) amplitude
is
calculated in the unitary \NNpiNN\ model, the initial and final nucleons are
dressed by pions and consequently each external nucleon obtains a
renormalization factor of $\tilde{Z}^{1/2}$.  The first renormalization problem
is the fact that $\tilde{Z}\ne Z$. This arises because two nucleons  cannot be
dressed at the same time in the truncated Hilbert space; thus, each nucleon in
a
two-nucleon state cannot obtain its full dressing. This, however, may not be
such a serious problem since, in practice, the difference between $Z$ and
$\tilde{Z}$ turns out to be quite small. The serious problem, instead, is the
size of the effective \piNN\ coupling constant in the \NNpiNN\ equations.
Taking
$Z\approx \tilde{Z}$, \fig{flaw}(b) illustrates that each vertex gets
renormalized by a factor of $Z$, so that the effective \piNN\ coupling constant
here becomes $Zf(m)$; this is a factor $Z^{1/2}$ times the physical coupling
constant. With $Z$ being typically between $0.6$ and $0.8$,  we come to the
disturbing conclusion that the effective \piNN\ coupling constant in the
\NNpiNN\ equations is smaller than the one used in constructing the input. This
observation helps explain why one typically obtains much too small \pppid\
cross
sections using this model [8-11]. 
\begin{figure}[t] \vspace{1.1cm}
\hspace{1cm} \special{illustration 1.ps}
\caption{\fign{flaw} Allowed dressing in the unitary \NNpiNN\ model, with
associated $Z$ renormalization factors. (a) \piN\ nucleon pole graph, (b) \NN\
OPE graph.} \end{figure}

These important observations about the renormalization problem in the unitary
\NNpiNN\ model were already made in 1985 by Sauer {\em et al.} \cite{Sauer},
yet they seem to have gone largely unnoticed. Perhaps this is partly because
one
could find less "fatal" reasons for the low cross sections; for example, it was
legitimately argued that off-shell effects and the lack of a "backward-going
pion" in the \NNND\ amplitude can lead to the underestimation of \pppid\ cross
sections \cite{Fayard}. However, with the advent of calculations where the
nucleon and $\Delta$ are treated on an equal footing in the \NNpiNN\ model
(the so-called \BBpiBB\ equations \cite{FewBody_Adelaide}), the effective
\piNN\
coupling constant is lowered by yet a further factor of $Z^{1/2}$ in the most
important \NNND\ amplitude, and it has become very apparent that the
renormalization problem is indeed fatal to this type of approach to the \piNN\
system.

It may seem that one can fix the renormalization problem "by hand" by
strategically including extra $Z^{1/2}$ factors in either \piNN\ propagators
or \piNN\ form factors, or both. But it soon becomes apparent that there is
no easy way of doing this without destroying the three-body unitarity
of the equations.

Here we report on a completely different formulation of the \piNN\ problem
where
unitary equations are obtained without having to truncate the Hilbert space to
some maximum number of pions. Consequently, all possible dressings of
one-particle propagators and vertices are retained in our model. In this way we
overcome the renormalization problems discussed above. The key element of our
derivation is the use of convolution integrals that have enabled us to sum all
the possible disconnected time-ordered graphs \cite{KB_PRC}. Indeed, we
basically make only one approximation in deriving the \piNN\ convolution
equations,  namely, we neglect connected three-body forces.

As the details of this new approach have already been published \cite{KB_PL},
here we shall give a shortened derivation, retaining only the essential steps.
A
substantial part of our presentation involves an extension of the convolution
idea to connected diagrams. In particular, we show how any diagram, connected
or
disconnected, can be calculated so that all possible dressing contributions are
included. We then use the derived formalism to calculate the \NN\ OPE potential
where the nucleons are fully dressed. Not only does this calculation indicate
that it may be important to include dressing into the popular one boson
exchange
models of the \NN\ interaction, but it also provides for us a first check on
the
size of the neglected three-body forces. We find that, in this case, the
three-body connected force contribution is negligible. \bigskip\medskip

\centerline{\bf THE CONVOLUTION \mbox{\boldmath $\pi N\!N$} EQUATIONS}
\medskip

We consider a time-ordered perturbation theory of nucleons and pions described
by a Hamiltonian $H$. The exact form of $H$ need not be specified.
The Green function for the \piNNpiNN\ process is thus defined by
\be
\la \p_1',\p_2',\p_3'| G(E) |\p_1,\p_2,\p_3\ra   =
\la \p_1',\p_2',\p_3'|\frac{1}{E^+ -H}|\p_1,\p_2,\p_3\ra    \eqn{G_piNN}
\ee
where $\p_\A$ ($\p_\A'$) denote initial (final) momenta; here, as below,
$\A=1,2$ label the two nucleons while $\A=3$ denotes the pion. Note that we
suppress spin-isospin labels in order to save on notation, and it is assumed
that the nucleons are distinguishable since antisymmetrization can be carried
out at the end.

We define the \piNNpiNN\ t-matrix $T(E)$ by the equation
\be
G(E) = G_0(E) + G_0(E) T(E) G_0(E)                  \eqn{T_def}
\ee
where $G_0(E)$ is the free \piNN\ propagator with all particles fully
dressed (i.e. it is the fully disconnected part of $G(E)$). Note that
in this section symbols $G$ and $g$ shall represent operators, while in the
rest of the paper they are usually numbers; this should, in any case, be clear
from the context. \eq{T_def} is thus a relation between operators acting in the
momentum space of one pion and two nucleons. In a previous work \cite{KB_PRC},
we have shown how $G_0(E)$ can be expressed in terms of the individual pion and
nucleon propagators by a convolution integral. In particular, if we write
\be
\la \p_\A'| g_{\A}(E) |\p_\A\ra    =
\la \p_\A'|\frac{1}{E^+ -H}|\p_\A\ra  \eqn{g_pi}
\ee
for the dressed nucleon and pion propagators, then $G_0(E)$ is given by
\be
G_0(E) = \left( - \frac{1}{2\pi i} \right)^2
\int_{-\infty}^{\infty} dz_1 dz_2 \,
g_1(E-z_1) g_2(z_1-z_2) g_3(z_2) .
\ee
Here it is understood that each single-particle propagator is an operator
acting in its own Hilbert space.

With $T(E)$ being defined as in \eq{T_def}, we can write
\be
T(E) = V(E) + V(E) G_0(E) T(E)    \eqn{LS}
\ee
where the potential $V(E)$ is the sum of all possible \piNN - irreducible
graphs excluding those consisting of fully disconnected \piNN\ states. At this
point we shall neglect the connected diagrams of potential $V(E)$; these
correspond to connected three-body forces and will be considered in the next
section. It should be stressed that the connected diagrams of $V$ are the only
diagrams that are neglected in this model.

For the following discussion we follow the labelling convention where
amplitudes that involve the pion interacting with nucleon $i$ are labelled by
subscript $i$, while the \NN\ interaction with a pion spectator is labelled by
subscript 3. Further, the indices $i$, $j$, are reserved for nucleons 1 and 2
while the Greek indices $\A$, $\B$, $\C$ go from 1 to 3 unless otherwise
indicated.
\begin{figure}[t]
\vspace{1.8cm}
\hspace{.5cm} \special{illustration 2.ps}
\vspace{-3mm}
\caption{\fign{V} Diagrammatic representation of potentials (a) $V_1$, (b)
$V_3$, and (c) $V_4$. The open circles represent all possible graphs excluding
those that lead to \piNN\ intermediate states. Potentials $V_2$ and $V_5$ are
obtained by interchanging the two nucleons.}
\end{figure}

Of special relevance to the following discussion is the work of Stelbovics and
Stingl (SS) \cite{SS}, who investigated the disconnectedness problem in the
\piNN\ system, and then applied it to a model where Hilbert space is chopped at
the two-pion level. As discussed by SS, all disconnected $3\rightarrow 3$
diagrams belong to one of five classes of disconnectedness, denoted by
$\delta_\A$, characterized by an appropriate momentum-space $\delta$-function.
Three of these classes, $\delta_1$, $\delta_2$, $\delta_3$, have two
of the particles interacting, the third being spectator. The class $\delta_4$
has the pion in initial state being absorbed by nucleon 2 with the final state
pion being emitted by nucleon 1. Class $\delta_5$ is the same as $\delta_4$ but
with nucleons 1 and 2 interchanged.

According to this classification, the potential $V(E)$ can be written as the
sum
\be
V(E) = \sum_{\A=1}^5 V_\A(E)
\ee
where $V_\A(E)$ consists of diagrams of disconnectedness $\delta_\A$.
The potentials $V_\A(E)$ are represented diagrammatically in \fig{V}. It is
worth mentioning that one of the contributions to the potential of \fig{V}(c)
is the so called Jennings term which has been shown to be important for
calculations of the tensor polarization $t_{20}$ in \pid\ scattering
\cite{Jennings}.

As it stands, \eq{LS} cannot be used directly for calculations as its kernel is
disconnected. To derive equations with a compact kernel we proceed by analogy
with the case of Faddeev equations and eliminate the potentials $V_\A(E)$ in
\eq{LS} in favour of completely summed contributions of disconnectedness
$\delta_\A$. Let us therefore denote by $\tilde{w}_\A$ the Green function
consisting of the set of {\em all} diagrams, reducible and irreducible,
belonging to the disconnectedness class $\delta_\A$. We can then define the
corresponding amplitudes $w_\A$ by
\be
\tilde{w}_\A = G_0 w_\A G_0 .  \eqn{tilde_w}
\ee

As $\sum_\alpha w_\alpha$ is the disconnected part of $T$, one can use \eq{LS}
to equate the disconnected parts of type $\delta_\alpha$, and thus obtain
relations between the $V_\alpha$ and the $w_\alpha$. These relations can then
be
used to reexpress \eq{LS} as \cite{SS}
\bea
T  &=& \sum_{\A=1}^5 T_\A  \\
T_\A &=& w_\A + \sum_{\B=1}^5 w_\A \K_{\A\B} G_0 T_\B      \eqn{SS}
\eea
where $\K_{\A\B}$ is a $5\times 5$ matrix with elements
$\K_{11}=\K_{22}=\K_{33}=\K_{51}=\K_{42}=\K_{14}=\K_{25}=\K_{45}=\K_{54}=0$,
all other elements being equal to $1$.
The first iteration of \eq{SS} then gives us a compact kernel.

As our equations and the ones of SS share the same disconnectedness structure,
it is especially interesting to compare the actual models used. In the case of
the SS model, the potentials $V_\A$ were specified by graphs of the lowest
possible order. It then turned out that their $w_\A$, needed to specify the
kernel of \eq{SS}, are not directly related to usual subsystem t-matrices, and
must therefore be specified by solving very complicated integral equations
\cite{AS}. By contrast, we take the most complicated possible choice for the
$V_\A$, namely the set of all possible contributing diagrams of field theory.
The remarkable thing is, that it is just this maximally complicated choice that
enables us to express the needed $w_\A$ as simple convolutions of dressed
subsystem amplitudes (the usual two-body t-matrices $t_\A$, the \piNN\ vertex
$f$, and the dressed one-nucleon propagator $g$) \cite{KB_PRC}. As these
convolution integrals are written for Green function quantities, we shall
utilize a "tilde" notation, as in \eq{tilde_w}, to label amplitudes with
additional initial and final-state propagators. Thus for the \piN\ t-matrix we
define \be \tilde{t}_i(E) = g_{\pi N_i}(E) t_i(E) g_{\pi N_i}(E)
\eqn{tilde_ti} \ee where $g_{\pi N}$ is the dressed \piN\ propagator. We note
further that  $\tilde{t}_i$ is the connected part of the full $\pi N_i$ Green
function, i.e.  \be
\la \p_i',\p_3'| \tilde{t}_i(E) |\p_i,\p_3\ra
\,\, = \,\, \la \p_i',\p_3| \frac{1}{E^+-H} |\p_i,\p_3\ra_c .
\ee

In a similar way we can define $\tilde{t}_3$.
With these definitions, $w_\A$ can be written directly in terms of
convolutions of $\tilde{t}_\A$ and the spectator particle propagators
$g_\A$. Introducing the short-hand notation
\be
c = a \otimes b
\ee
to mean the convolution integral
\be
c(E) = - \frac{1}{2\pi i} \int_{-\infty}^{\infty} dz\, a(E-z) b(z) ,
\ee
we have that ($i\ne j$)
\bea
\begin{array}{ccc}
\tilde{w}_i = \tilde{t}_i \otimes g_j & ; &
\tilde{w}_3 = \tilde{t}_3 \otimes g_3.
\end{array}
\eea

It should be noted that $t_i(E)$ is the full \piN\ t-matrix
and can therefore be written in terms of nucleon pole and non-pole parts:
\be
t_i(E) = f_i(E) g_i(E) \bar{f}_i(E) + t_i^b(E)      \eqn{t_i}
\ee
where $t_i^b(E)$ is the non-pole part, $f_i(E)$ is the dressed vertex for $
N\!\rightarrow \!\pi N$, and $\bar{f}_i(E)$ is the dressed vertex for
$\pi N\!\rightarrow\!N$. These vertices are explicitly given in terms of
$\tilde{f}_i$ and $\tb{f}_i$:
\be
\begin{array}{ccc}
\tilde{f}_i = g_{\pi N_i} f_i g_i & ; &
\tb{f}_i = g_i \bar{f}_i g_{\pi N_i}
\end{array}
\ee
where
\bea
\la\p_i',\p_3'| \tilde{f}_i(E) |\p_i\ra &=&
\la\p_i',\p_3'| \frac{1}{E^+-H} |\p_i\ra \\
\la\p_i'| \tb{f}_i(E) |\p_i,\p_3\ra &=& \la\p_i'| \frac{1}{E^+-H} |\p_i,\p_3\ra
{}.
\eea

In a similar way, we can also express $w_4$ and
$w_5$ directly in terms of the dressed \piNN\ vertex using
\be
\begin{array}{lll}
\tilde{w}_4 = \tilde{f}_1 \otimes \tb{f}_2 & ; &
\tilde{w}_5 = \tilde{f}_2 \otimes \tb{f}_1 .
\end{array}                              \eqn{V_4^R_conv}
\ee

As the \NN\ channel is hidden in the input terms $w_\A$, the
relation of \eq{SS} to the full \NN\ amplitude is not clear. Neither
is it apparent if in the process of iteration, the intermediate \NN\
propagators will obtain their full dressing. For these reasons, we would like
to recast \eq{SS} into a form that explicitly exposes the \NN\
channel.

For this purpose, we write \eq{SS} in an AGS form \cite{SS}
\be
U_{\A\B}= \kappa_{\A\B} G_0^{-1} + \sum_{\C=1}^5 \kappa_{\A\C} w_\C G_0
U_{\C\B} .                                       \eqn{U}
\ee
We note, however, that in the $\kappa$-matrix, rows 1 and 5
are identical, as are rows 2 and 4, columns 1 and 4, and columns 2 and 5.
This means that
\be
\begin{array}{lllllll}
U_{\A 4} = U_{\A 1} & ; &  U_{\A 5} = U_{\A 2} & ; &
U_{4 \B} = U_{2 \B} & ; &  U_{5 \B} = U_{1 \B} . \end{array} \eqn{U_symmetry}
\ee
Thus although \eq{U} is a set of $5\times 5$ coupled equations, it can be
reduced to a set of $3\times 3$ equations.

We recall that the terms $\tilde{w}_i$ $(i=1,2)$ are defined as
the set of all possible disconnected graphs for pion scattering on nucleon $i$.
In order to expose two-nucleon states, we define the Green function
$\tilde{w}_i^P$ as the set of all two-nucleon reducible graphs belonging to
$\tilde{w}_i$, and write for the corresponding amplitudes
\be
w_i = w_i^0 + w_i^P      \eqn{w_i}
\ee
where $w_i^0$ is two-nucleon irreducible. Since we consider all possible
contributions, it is clear that one can write
\be
w_i^P = F_i G_{N\!N} \bar{F}_i     \eqn{w_i^P}
\ee
where $G_{N\!N}$ is the fully dressed two-nucleon propagator, $F_i$ and
$\bar{F}_i$, are the
fully dressed \piNN\ vertices in the two-nucleon sector. Again we can write
these in terms of the one-nucleon propagator $g_i$ and \piNN\ vertices $f_i$
and $\bar{f}_i$ through the convolution expressions $(i\ne j)$
\bea
G_{N\!N}    &=&         g_1 \otimes g_2 ,     \eqn{G_NN_conv}   \\
G_0 F_i G_{N\!N}  &=& \tilde{f}_i \otimes g_j ,     \eqn{tilde_F_i_conv}\\
G_{N\!N}  \bar{F}_i G_0 &=&    \tb{f}_i \otimes g_j .     \eqn{tb_F_i_conv}
\eea
In the same way we can separate out the two-nucleon reducible contributions
from $w_4$ and $w_5$, and with the further convention that $w_3^P=0$, we can
write generally
\be
w_\A = w_\A^0 + w_\A^P .       \eqn{w_A}
\ee
Clearly
\bea
\begin{array}{ccc}
w_4^P = F_1 G_{N\!N} \bar{F}_2    & ; &
w_5^P = F_2 G_{N\!N} \bar{F}_1 .  \eqn{w_4^P}
\end{array}
\eea
With these definitions we can define the amplitudes $U_{\A\B}^0$ corresponding
to all possible two-nucleon irreducible contributions, namely
\be
U_{\A\B}^0 = \kappa_{\A\B} G_0^{-1}
            + \sum_{\C=1}^5 \kappa_{\A\C} w_\C^0 G_0 U_{\C\B}^0 .   \eqn{U^0}
\ee
{}From \eq{U}, \eq{w_A} and \eq{U^0}, it follows that
\be
U_{\A\B} = U_{\A\B}^0
+ \sum_{\C=1}^5 U_{\A\C}^0 G_0 w_\C^P G_0 U_{\C\B}.   \eqn{U=U^0}
\ee
Using the explicit forms for $w_\C^P$ given in \eq{w_i^P} and \eqs{w_4^P},
and making use of \eq{U_symmetry} to eliminate channels 4 and 5,
we obtain for the last term in \eq{U=U^0}
\be
\sum_{\C=1}^5 U_{\A\C}^0 G_0 w_\C^P G_0 U_{\C\B} =
(U_{\A 1}^0 G_0 F_1 + U_{\A 2}^0 G_0 F_2) \, G_{N\!N} \,
(\bar{F}_1 G_0 U_{1\B} + \bar{F}_2 G_0 U_{2\B}).         \eqn{factorisation_1}
\ee
Substituting into \eq{U=U^0}
leads directly to the equation for nucleon-nucleon scattering
\be
T_{N\!N}(E) = V_{N\!N}(E) + V_{N\!N}(E) G_{N\!N}(E) T_{N\!N}(E)
\ee
where
\bea
V_{N\!N}(E) &=& \sum_{ij} \bar{F}_i G_0 U_{ij}^0 G_0 F_j  \eqn{V_NN} \\
T_{N\!N}(E) &=& \sum_{ij} \bar{F}_i G_0 U_{ij} G_0 F_j .  \eqn{T_NN}
\eea
This derivation shows explicitly that the \NN\ propagator, which we did not
have as an explicit input in \eq{U}, is indeed the fully dressed $G_{N\!N}$
which is given in terms of the convolution integral of \eq{G_NN_conv}. Although
it may be clear that the amplitude $U_{ij}$ is made up of fully dressed
component amplitudes, it is noteworthy that the connection to the \NN\
scattering amplitude is via the vertex functions $F_i$ and $\bar{F}_i$ which
themselves have all possible dressings.

A feature of our formulation is that the input to our derived equations,
\eq{SS} or equivalently \eq{U}, consists just of usual two-body subsystem
off-shell amplitudes. This is an important aspect for practical calculations
and was not the case for the formally similar equations of SS.
On the other hand, due to the terms $w_4^0$ and $w_5^0$, the kernel of the
equations nevertheless is not of the standard type where all interactions are
of
two-body type. We shall now show how one can rewrite the equations to have
standard pair-like interactions in the kernel but at the expense of introducing
an extra dimension in the \NN\ sector.

We start by writing an alternative decomposition of $w_\A$ to that given in
\eq{w_A}. For $w_i$ we convolute $t_i$, as given in \eq{t_i}, with a dressed
spectator nucleon; that is, the decomposition
\be
w_i = \bar{w}_i^0 + \bar{w}_i^P      \eqn{bar{w}}
\ee
will be defined via the convolutions
\bea
G_0 \bar{w}_i^P G_0 &=& (g_{\pi N} f g \bar{f} g_{\pi N})_i \otimes g_j
                                                \eqn{bar{w}_i^P}  \\
G_0 \bar{w}_i^0 G_0 &=& (g_{\pi N} t^b g_{\pi N})_i \otimes g_j .
                                                \eqn{bar{w}_i^0}
\eea
We can now write $w_\A = \bar{w}_\A^0 + \bar{w}_\A^P$ generally by defining
$\bar{w}_3^P = 0$, $\bar{w}_4^P = w_4$, $\bar{w}_5^P = w_5$,
$\bar{w}_3^0 = w_3$, $\bar{w}_4^0 = 0$, and $\bar{w}_5^0 = 0$.
Analogous to \eq{bar{w}_i^P} we have that
\bea
G_0 \bar{w}_4^P G_0 =
(g_{\pi N} f  g)_1 \otimes (g \bar{f} g_{\pi N})_2, \eqn{bar{w}_4^P}     \\
G_0 \bar{w}_5^P G_0 =
(g_{\pi N} f g)_2 \otimes (g \bar{f} g_{\pi N})_1,  \eqn{bar{w}_5^P}
\eea
which are just more explicit versions of \eq{V_4^R_conv}.

Introducing this decomposition of $w_\A$ into \eq{U}, leads us to define,
by analogy to \eq{U^0},
\be
\bar{U}_{\A\B}^0 = \kappa_{\A\B} G_0^{-1}
            + \sum_{\C=1}^5 \kappa_{\A\C} \bar{w}_\C^0 G_0 \bar{U}_{\C\B}^0 .
                                \eqn{bar{U}^0}
\ee
Analogous expressions to \eq{U=U^0} and \eq{factorisation_1} are also easily
written. Using a
short-hand notation, illustrated generally by
\bea
\begin{array}{lll}
(ab)_{1z} = a_1(E-z) b_1(E-z)  &;&  (ab)_{2z} = a_2(z) b_2(z) ,
\end{array}
\eea
we are then led to define a new kind of \NN\ potential and t-matrix depending
on
extra variables:
\bea
V_{N\!N}(z',z;E) &=&
\sum_{ij} (\bar{f}g_{\pi N})_{iz'} \bar{U}_{ij}^0
(g_{\pi N}f)_{jz}\eqn{V_NN_z}\\
T_{N\!N}(z',z;E) &=&
\sum_{ij} (\bar{f}g_{\pi N})_{iz'} U_{ij} (g_{\pi N}f)_{jz} .      \eqn{T_NN_z}
\eea
In this way the integral equation for \NN\ scattering acquires an extra
dimension,
\bea
\lefteqn{
T_{N\!N}(z',z;E) = V_{N\!N}(z',z;E) } \nn
& &- \frac{1}{2\pi i} \int_{-\infty}^{\infty} dz''\,
V_{N\!N}(z',z'';E) g_1(E-z'') g_2(z'') T_{N\!N}(z'',z;E) .    \eqn{T_NN_4D}
\eea
How $T_{N\!N}(z',z;E)$ can be identified with the physical \NN\ t-matrix,
$T_{N\!N}(E)$, can be seen as follows.
{}From \eq{tilde_F_i_conv}, it is easy to show
that in the case when the two initial-state nucleons are on-energy-shell,
i.e. $\bfp_1^2/2m+\bfp_2^2/2m+2m=E$, then
\be
g_{\pi N_i}(\bfp_i^2/2m+m)f_i(\bfp_i^2/2m+m)|\bfp_1\bfp_2\ra
= G_0(E)F_i(E)|\bfp_1\bfp_2\ra \eqn{gf}
\ee
with a similar equation holding for $\bar{f}_i g_{\pi N_i}$.
It can therefore be seen from \eq{T_NN_z} that if the initial and
final nucleons are on-energy-shell, and if we set the $z$ variables to the
energies of the second nucleon, i.e.
$z'=\bfp_2'\,^2/2m+m$, $z=\bfp_2^2/2m+m$, then
\be
\la\bfp_1'\bfp_2'|T_{N\!N}(z',z;E)|\bfp_1\bfp_2\ra  =
  \la\bfp_1'\bfp_2'|T_{N\!N}(E)|\bfp_1\bfp_2\ra
\ee
where $T_{N\!N}(E)$ is given in \eq{T_NN}.
In this sense $T_{N\!N}(z',z;E)$ can be
considered as an off-shell scattering amplitude with
additional energy-like variables $z$ and $z'$.

Our final result for the description of the \piNN\ system consists of the
coupled equations (\ref{bar{U}^0}) and (\ref{T_NN_4D}). The equation for
$\bar{U}^0$ is just the Faddeev equation for the \piNN\ system with no
absorption. In momentum space it is a 6-dimensional integral equation
and involves pair-interactions $\pi$-$N$ and $N$-$N$.

The equation for \NN\ scattering, \eq{T_NN_4D}, in
momentum space,  is a 4-dimensional integral equation. In this sense it is
similar to the Bethe-Salpeter equation but with particular forms for the
nucleon
propagator and \NN\ potential. In the case of \eq{T_NN_4D}, the nucleon
propagator contains only positive energy states while the one-pion exchange
potential, for example, has \piNN\ vertex functions that depend only on one
energy variable - in the Bethe-Salpeter case they depend on two energy
variables. It is interesting to note that by introducing a fourth dimension
into the \NN\ sector, we take into account the \NN\ irreducible diagrams in
$w_4$ and $w_5$ while at the same time having only pair-like interactions
in the kernel for $\bar{U}^0$.

We conclude by noting that our derivation is model independent. Thus,
for example, there was no need to specify either how pions actually couple to
nucleons, or what model is used for dressing the pion. We also note that our
expressions for the coupled \piNN\ equations, \eq{bar{U}^0} and \eq{T_NN_4D},
or alternatively the coupled equations (\ref{U^0}) and (\ref{T_NN}), express
the main features of the physics. Thus, for example, we can easily check that
our formulation gives consistent renormalization, simply by substituting the
pole term $Z/(E^+-m)$ wherever the single-nucleon propagator $g(E)$ appears.
The unitarity of our equations is also evident from their derivation. For
practical calculations, however, either of the coupled set of equations can
easily be cast into one set of coupled equations for the reactions \NNNN ,
\pidpid , and \NNpid . In the case of \eq{bar{U}^0} and \eq{T_NN_4D}, this
would result in coupled equations that are of similar form to the well known
unitary \NNpiNN\ equations \cite{AB_80}, but with an extra dimension in the
\NN\
sector.
\bigskip\bigskip

\centerline{\bf THE DRESSED \mbox{\boldmath $N\!N$} OPE POTENTIAL}
\bigskip

The essential feature of the \piNN\ equations, presented above, is that they
include all possible disconnected \piNN -irreducible graphs. This is the
first time that {\em all} such disconnected diagrams have been included, and
has
been achieved only after preliminary work showing that convolution integrals
can
sum all possible time orderings of {\em disconnected} graphs of time-ordered
perturbation theory \cite{KB_PRC}.

With all the disconnected diagrams included, the focus of attention turns
naturally to the question of how large the neglected {\em connected}
\piNN -irreducible graphs are. This is not an easy question to answer without
a systematic analysis of the whole range of such connected three-body
force contributions. Here we shall take the first steps in such an analysis
by examining the most basic ingredient of the \NN\ force, namely, the \NN\
one pion exchange potential.

Within the formalism of the \piNN\ convolution equations, the OPE potential
takes the form appearing in \eq{V_NN} with $U_{ij}^0$ replaced by
$G_0^{-1}$ (the inhomogeneous term of \eq{U^0}). It will be sufficient to
examine just one time ordering of the exchanged pion, and in this case, the OPE
potential of the \piNN\ convolution equations is given by
\be
V_{12}^{OPE} = \bar{F}_1 G_0 F_2 .          \eqn{V_12}
\ee

The consequence of neglecting connected three-body forces becomes immediately
apparent upon examination of the dressing contributions to this OPE potential.
In \eq{V_12}, each of the terms $\bar{F}_1$,  $G_0$, and $F_2$, is by itself
completely dressed. Thus a simple example of a dressing contribution that is
included in \eq{V_12} is given by \fig{OPE}(a). On the other hand, the
topologically similar graph of \fig{OPE}(b) is not included in the formalism of
the previous section since it is of the form $\bar{F}_2 V^c F_1$ where $V^c$ is
the connected \piNNpiNN\ three-body force.

In order to see the effect of neglecting dressings of the OPE potential
like that of \fig{OPE}(b), it would be useful to have an expression for the
fully dressed OPE potential. That is, we need the OPE graph where all possible
dressing graphs involving a \piNN\ vertex are retained (the pion itself,
however, will be assumed to have no dressing contributions). As will be shown
in
the sections below, such an expression can indeed be easily derived by
extending
the convolution idea to connected diagrams. Here we just give the result for
the corresponding OPE Green function (in the c.m.\ system) with momentum
conserving $\delta$-functions removed:
\bea
\lefteqn{\tilde{V}_{12}^{OPE}(\p',\p;E) = \left(-\frac{1}{2\pi i}\right)^2 \int
dz\, dz'\, g(z',\p') \bar{f}(\p',\p,z',z) g(z,\p) \frac{1}{z'-z-\omega_k} }
\hspace{4cm} \nn
&&g(E-z',-\p') f(-\p',-\p,E-z',E-z) g(E-z,-\p) .  \hspace{1cm}
\eqn{tilde_V_OPE}
\eea
\begin{figure}[t]
\vspace{1.8cm}
\hspace{2cm}\special{illustration 3.ps}
\caption{\fign{OPE}
(a) Example of a dressing diagram included in the convolution \piNN\
equations. (b) Example of a dressing diagram involving a connected three-body
force - such diagrams are not included in the convolution \piNN\ equations.}
\end{figure}
\noindent
Here $\omega_k$ is the energy of the exchanged pion, and $g(E,\p)$ is the
dressed nucleon propagator defined as the matrix element of \eq{g_pi}, but with
the momentum conserving $\delta$-function removed, i.e. \be
\delta(\p-\p') g(E,\p) = \la \p'|\frac{1}{E^+-H}|\p\ra .  \eqn{g}
\ee

A novel feature of \eq{tilde_V_OPE} is the need for a \piNN\ vertex function
$f(\p',\p,E',E)$ that depends on two energy variables. By contrast, in the
\piNN\ convolution equations, as in most other formulations utilizing
time-ordered perturbation theory, the dressed vertex depends only on one energy
variable. The relation between one- and two-energy vertices will be given
below.

The essential point here is that if one neglects connected three-body forces
from  \eq{tilde_V_OPE}, then one will obtain the OPE Green function of the
convolution \piNN\ equations, which numerically is given as a product of three
convolution integrals:
\bea \lefteqn{\tilde{V}_{12}^{OPE}(\p',\p;E) =
\left(-\frac{1}{2\pi i}\right)^3 \left[ \int dz\, g(z,\p') \bar{f}(\p',\p,z)
g(z,\p) g(E-z,-\p') \right] } \nn &&\left[ \int dz\, g(z,\p)
g(E-\omega_k-z,-\p') \right] \left[ \int dz\, g(z,-\p') f(-\p',-\p,z) g(z,-\p)
g(E-z,\p) \right] . \nn && \eqn{tilde_V_OPE_1}
\eea

The fully off-shell OPE potential $V^{OPE}_{12}(\p',\p;E)$ is related to the
OPE
Green function in the usual way:
\be
V^{OPE}_{12}(\p',\p;E) = G_{N\!N}^{-1}(E,\p',-\p')
\tilde{V}_{12}^{OPE}(\p',\p;E)
G_{N\!N}^{-1}(E,\p,-\p)    \eqn{fully-off-shell_V}
\ee
where $G_{N\!N}$ is the dressed two-nucleon propagator given as in
\eq{G_NN_conv} .
\begin{figure}[t] \vspace{6cm}
\hspace{.5cm}\special{illustration 4.ps} \caption{\fign{comparison}
Comparison of the half-off-shell dressed $N\!N$ OPE potential,
$V^{OPE}_{12}(\p_0,\p;E)$ calculated with full dressing (solid curves), and
within the convolution \piNN\ model (long-dashed curves). The short-dashed
curve
is for the case where no dressing is included.} \end{figure}
For the numerical comparison of \eqs{tilde_V_OPE} and (\ref{tilde_V_OPE_1}) we
follow our previous work \cite{KB_PRC} and use the $M1$ \piN\ interaction of
Ref.\ \cite{AM}. We shall, in each case, calculate the half-off-shell potential
$V^{OPE}_{12}(\p_0,\p;E)$, defined by \eq{fully-off-shell_V} in the limiting
case $\p'\rightarrow \p_0$, where $\p_0$ is the on-shell momentum, i.e.
$E=p_0^2/m+2m$. Here we shall consider the simplified case where no dressing is
included for the vertices of the  exchanged pion; in this case, both the
two-energy vertex $f(\p',\p,E',E)$ of \eq{tilde_V_OPE} and the single-energy
vertex $f(\p',\p,E)$ of \eq{tilde_V_OPE_1}, reduce down to the
energy-independent bare vertex $f_0(\p',\p)$. Note that the restriction to bare
vertices is limited to the pion exchanged between the two nucleons, so that the
dressed nucleon propagators $g$ are totally unaffected by this simplification.
The half-off-shell potential $V^{OPE}_{12}(\p_0,\p;E)$ depends on three
variables, the energy $E$, the magnitude $p$ of the off-shell momentum, and the
cosine of the angle between $\p_0$ and $\p$. For the  numerical comparison, we
have examined the potential as a function of energy $E$ for a large range of
values of $p$ and  $x=\hat{\p}_0\cdot\hat{\p}$. A typical result is shown in
\fig{comparison} where, in this case, we have set $p=.1 p_0$ and $x=-0.8$. In
this figure, the solid curve gives the OPE potential with full dressing, while
the long-dashed curve is the result when connected three-body forces are
neglected. Also shown is the standard OPE potential where no dressing at all is
included (short-dashes). The first observation of note is the significant
effect
that dressing has on the standard OPE potential. This raises questions about
the
role such dressing may play in standard descriptions of the \NN\ force. The
second observation forms the main result of this section -  it is the essential
identity of the solid and long-dashed curves, indicating that the contribution
of connected three-body forces to the OPE potential is negligible. Although
strictly applying to the OPE potential with bare vertices, this is a very
encouraging  result that gives us hope that connected three-body forces may be
small in general. If this is borne out in further studies, this would mean
that,
for the \piNN\ system, the convolution \piNN\ equations provide a way of
effectively summing all possible diagrams of time-ordered perturbation theory.
\bigskip\bigskip

\centerline{\bf DRESSING OF CONNECTED DIAGRAMS - METHOD I}
\medskip
In Ref.\ \cite{KB_PRC} we showed how a convolution integral can sum all
possible
relative time orderings of a {\em disconnected} diagram of time-ordered
perturbation theory. This was then used to demonstrate how two disconnected
nucleons can be dressed simultaneously. Here we show that the convolution idea
can also be easily extended to {\em connected} diagrams. The final result shows
that the sum of all topologically similar diagrams, connected or disconnected,
differing only in the relative time ordering of their vertices, can be
expressed
through a convolution formula that effectively integrates out initial and final
relative energies from the topologically equivalent Feynman diagram. One
application of this generalized convolution formula is to the problem of
including all possible dressings in connected diagrams.

The starting point of the following discussion is relativistic quantum field
theory. Since Feynman diagrams contain, in some sense, all time orderings, it
is
not surprising that our goal will be to show how to extract the Green
function of time-ordered perturbation theory from the Green function of
relativistic quantum field theory. Although the procedure we'd like to follow
is
general, for presentation purposes we specifically consider the  \NNNN\ process
where the interaction is described by a Hamiltonian $H=H_0+H_I$ involving meson
and baryon fields. The explicit form of $H$ need not be specified.

The free fermion field at time $t=0$ is denoted by $\psi(\x)$. At time $t$, we
define
\bea
&&\psi(\x,t) = e^{iH_0t} \psi(\x) e^{-iH_0t}    \eqn{psi}  \\
&&\Psi(\x,t) = e^{iHt} \psi(\x) e^{-iHt}        \eqn{Psi}
\eea
being the interaction picture and Heisenberg fields, respectively. The free
fermion field $\psi(x)=\psi(\x,t)$ can then be written in terms of its Fourier
decomposition as \cite{BD}
\be \psi(x) = \sum_{s} \int
\frac{d^3p}{(2\pi)^{3/2}}\sqrt{\frac{m}{E_p}} [b(\p,s)u(\p,s)
e^{-ip\cdot x} + d^\dagger(\p,s)v(\p,s) e^{ip\cdot x} ].
\eqn{free_fermion}    \ee

We consider the \NNNN\ process described by the coordinate space Green function
\be
i\G(x,y;x',y') = \laa 0| T \Psi(x) \Psi(y) \Psib(x') \Psib(y') |0\raa  ,
                                                        \eqn{GPsi}
\ee
which, because each $\Psi$ is a four-component spinor, can be considered as
a  $16\times 16$ matrix. Here $|0\raa$ is the dressed vacuum, its relation to
the
bare vacuum $|0\ra$ being given by
\be
|0\raa = \frac{1}{\laa 0| 0\ra} [1+\frac{1}{0^+-H}H_I]|0\ra  .
\eqn{dressed_vacuum}
\ee
The momentum space Green function is defined by
\be
\G(p,q;p',q') = \int e^{i(x\cdot p + y\cdot q -x'\cdot p' - y'\cdot q')}
\G(x,y;x',y') d^4x\,d^4y\,d^4x'\,d^4y' ,     \eqn{FT}
\ee
which can also be expressed without the momentum conserving $\delta$-function
by defining the Green function $G(p,q;p',q')$:
\be
\G(p,q;p',q') = (2\pi)^4 \delta^4(p'+q'-p-q) G(p,q;p',q') .
\ee

We follow Logunov and Tavkhelidze \cite{LT} and consider the two-time Green
function
\be
\G(\x,\y,t;\x',\y',t') \equiv \G(x,y;x'y'){\scriptsize\left|_{ \begin{array}{l}
x_0=y_0=t\\x'_0=y'_0=t'\end{array} } \right. }  .    \eqn{Gtt_x}
\ee
The two-time Green function in momentum space is defined by
\bea
\lefteqn{ \tilde{\G}(\p,\q,E;\p',\q',E') \equiv \int e^{i(-\x\cdot \p - \y\cdot
\q +\x'\cdot \p' + \y'\cdot \q' + tE - t'E')} }\hspace{5cm} \nn
& &\G(\x,\y,t;\x',\y',t') d^3x\,d^3y\,d^3x'\,d^3y'\,dt\, dt' .\hspace{3cm}
\eqn{Gp}
\eea
A little  algebra shows that
\be
\tilde{\G}(\p,\q,E;\p',\q',E') = \frac{1}{(2\pi)^2}\int_{-\infty}^\infty
d\omega\int_{-\infty}^{\infty}d\omega'\,
\G(\p,E-\omega,\q,\omega;\p',E'-\omega',\q',\omega').   \eqn{G_convo_delta}
\ee
This result was used by Logunov and Tavkhelidze as the starting point for their
study of the two-time Green function. It also constitutes the solution to our
problem as we now proceed to show.

It is useful to express $\tilde{\G}$ without momentum conserving
$\delta$-functions, thus we define the Green function $\tilde{G}$ by
\be
\tilde{\G}(\p,\q,E;\p',\q',E') = (2\pi)^4\delta(\p+\q-\p'-\q')\delta(E-E')
\tilde{G}(E,\p,\q;\p',\q').             \eqn{G_nodelta}
\ee
Removing the momentum conserving $\delta$-functions from \eq{G_convo_delta},
gives  \be
\tilde{G}(E,\p,\q;\p',\q') = \frac{1}{(2\pi)^2}\int_{-\infty}^\infty
d\omega\int_{-\infty}^{\infty} d\omega'\,
G(\p,E-\omega,\q,\omega;\p',E-\omega',\q',\omega').   \eqn{G_convo}
\ee

The rhs of \eq{G_convo} is a double convolution over initial and final relative
energies of Feynman graphs. The goal, therefore, is to show how the lhs,
$\tilde{G}(E,\p,\q;\p',\q')$, is related to the Green function of time-ordered
perturbation theory. For this we go back to coordinate space and examine the
two-time Green function in detail. By the definition of the time-ordered
product in \eq{GPsi},
\bea
i\G(\x,\y,t;\x',\y',t')
&=& \theta(t-t')\laa 0|\Psi(\x,t)\Psi(\y,t)\Psib(\x',t')\Psib(\y',t')|0\raa \nn
&+& \theta(t'-t)\laa 0|\Psib(\x',t')\Psib(\y',t')\Psi(\x,t)\Psi(\y,t)|0\raa .
\eqn{ret_adv}
\eea
Using \eq{Psi} in this equation, and taking the Fourier transform as in \eq{Gp}
gives the result
\bea
(2\pi)^3\delta(\p+\q-\p'-\q')\tilde{G}(E,\p,\q;\p',\q') \! &=& \! \laa
0|\psi(\p)\psi(\q)\frac{1}{E^{+}-H} \psib(\p')\psib(\q')|0\raa  \nn
&-&\! \laa 0|\psib(\p')\psib(\q')\frac{1}{E^{-}+H} \psi(\p)\psi(\q)|0\raa
\eqn{topt_1} \hspace{1cm}
\eea
where we have used \eq{G_nodelta}, the fact that
\be
\int_{-\infty}^{\infty} \theta(t) e^{i\omega t}dt = \frac{i}{\omega+i\epsilon},
\eqn{good_trick}
\ee
and where we have introduced the momentum space fields
\be
\psi(\p) \equiv \int d\x^3\,e^{-i\x\cdot\p}\psi(\x) .   \eqn{psip_def}
\ee
Using \eq{free_fermion} at $t=0$ in \eq{psip_def}, one obtains that
\be
\psi(\p) = \sum_{s} \sqrt{\frac{(2\pi)^3m}{E_p}} [b(\p,s)u(\p,s) +
d^\dagger(-\p,s)v(-\p,s) ]. \eqn{psip}
\ee
\eq{topt_1} is basically the result that we seek. It expresses
$\tilde{G}$ in terms of two terms corresponding to the retarded and advanced
parts of the Green function of \eq{ret_adv}; we shall correspondingly
name the first and second terms on the rhs of \eq{topt_1} as retarded and
advanced, respectively.

One may compare \eq{topt_1} with the Green function of time-ordered
perturbation theory, which for the process in question is given by
\be
(2\pi)^3\delta(\p+\q-\p'-\q') G_{\alpha\beta,\alpha'\beta'}(E,\p,\q;\p',\q') =
\la \p,\alpha,\q,\beta|\frac{1}{E^{+}-H} |\p',\alpha',\q',\beta'\ra  \eqn{Gto}
\ee
where the spin components $\alpha,\beta,\alpha',\beta'$ are shown explicitly.
Taking into account \eq{psip}, we see that the retarded part of $\tilde{G}$
in \eq{topt_1} is very similar to the Green function of \eq{Gto}. This
similarity suggests a simple transformation of \eq{topt_1} defined by the
equation
\be
G_{\alpha\beta,\alpha'\beta'} =
\left[  \frac{1}{(2\pi)^3
m}\sqrt{E_{p}E_{q}}\bar{u}(\p,\alpha)\bar{u}(\q,\beta)
\right] G \left[
 u(\p',\alpha')u(\q',\beta') \frac{1}{(2\pi)^3 m}\sqrt{E_{p'}E_{q'}} \right]
\eqn{transformation}
\ee
where here $G=G(\p,\q,\p',\q')$ represents any appropriate Green function.
Note that \eq{transformation} transforms a $16\times 16$ matrix in spinor space
to a $4\times 4$ matrix in spin space. Under this transformation, \eq{topt_1}
becomes \bea \lefteqn{ (2\pi)^3
\delta(\p+\q-\p'-\q') \tilde{G}_{\alpha\beta,\alpha'\beta'}(E,\p,\q;\p',\q') =}
\hspace{4cm}\nn
&& \laa 0|b(\p,\alpha)b(\q,\beta)\frac{1}{E^+-H}b^\dagger(\p',\alpha')
b^\dagger(\q',\beta')|0\raa  \nn
&-& \laa 0|b^\dagger(\p',\alpha') b^\dagger(\q',\beta')\frac{1}{E^-+H}
b(\p,\alpha)b(\q,\beta)|0\raa . \hspace{2cm} \eqn{G_tilde_transformed} \eea

If $|0\raa = |0\ra$, the advanced term of \eq{G_tilde_transformed} disappears,
and the retarded term becomes identical with the Green function of time ordered
perturbation theory, \eq{Gto}. Thus the only difference between the transformed
two-time Green function of \eq{G_tilde_transformed}  and the one of
time-ordered perturbation theory, is in the type of vacua used to take the
matrix element: the transformed version of $\tilde{G}$ uses the dressed vacuum
while standard time-ordered perturbation theory, as in \eq{Gto}, uses the bare
vacuum.

In general, $|0\raa \ne |0\ra$, however, in the special case where the
dressed and bare vacua are equal, \eqs{Gto} and (\ref{G_tilde_transformed})
become identical. In this case, \eq{G_convo}, after the transformation of
\eq{transformation}, provides us with a formula that expresses
the full Green function of time-ordered perturbation theory in terms of a
double convolution integral of the full Feynman Green function. This result
can be easily generalized to hold for perturbation diagrams of a given
order in the interaction, or even for perturbation diagrams of a given topology
(the argument is similar to the one used in Ref.\ \cite{KB_PRC}).
It is in this latter form that the method can give particularly useful
formulas,
like that of \eq{tilde_V_OPE}, where all possible time-orderings of a
perturbation graph
are summed by performing convolution integrals of one Feynman diagram.

A similar result to ours has lately been obtained  by Phillips
and Afnan \cite{Daniel} using a much more involved argument. In their case,
however, they neglected antinucleons altogether in order to obtain the
connection of \eq{G_convo} to time-ordered perturbation theory.\footnote{This
paper claims, incorrectly, to have shown that the convolution formula, derived
by us in Ref.\ \cite{KB_PRC}, is a special case of their two-time approach
where
antinucleons are neglected. If anything, just the opposite is the case: our
convolution formula was derived using a method, akin to Method II, which does
not need for its validity any approximations whatsoever. We also do not share
these authors' criticism of three-dimensional approaches, even though we
have ourselves already presented the first consistent four-dimensional approach
\cite{KB_NP}.}

We emphasize that all that is needed in our derivation is the condition
$|0\raa = |0\ra$. For this to be true, it is sufficient to drop terms
from the interaction Hamiltonian that connect vacuum to vacuum states,
otherwise antinucleons can be retained. For example, in the simple case of
interaction  $H_I \sim \psib\phi\psi$, expanding each field in terms of
creation
and  annihilation operators gives eight terms; however, of these eight, only
terms $b^\dagger a^\dagger d^\dagger$ and $d a b$ contribute to the dressing of
the vacuum (in lowest order via the process $0\rightarrow
N\!\bar{N}\pi\rightarrow 0$), and just these terms can be dropped while still
retaining other terms involving $d$ and $d^\dagger$.
\bigskip\bigskip

\centerline{\bf DRESSING OF CONNECTED DIAGRAMS - METHOD II}
\medskip

The above method using the two-time Green function can be used to derive
\eq{tilde_V_OPE} for the dressed \NN\ OPE potential, but only in the case where
vacuum dressing has been neglected. It turns out, however, that
\eq{tilde_V_OPE}
is more general, holding for {\em any} Hamiltonian $H$, including ones
involving
vacuum dressing.

Here we derive \eq{tilde_V_OPE} using a totally different method that is close
in  spirit to the one used to derive the convolution formula for disconnected
diagrams \cite{KB_PRC}. This method involves the temporary introduction
of different types of pions and nucleons into a time-ordered perturbation
theory description, and provides a way to derive \eq{tilde_V_OPE} without
making
any approximations. In this sense, this method is more general than the one
based on the two-time Green function.

The central idea is to interpret the dressed \NN\ OPE potential as
consisting of two distinguishable nucleons $N_1$ and $N_2$, each being dressed
by its own pion, $\pi_1$ and $\pi_2$ respectively, and in addition, exchanging
a
third type of pion $\pi$; of the three types of pion, only $\pi$ can
interact with both nucleons. We accordingly define $H_1 = H_0(1) + H_I(1)$ to
be the Hamiltonian describing the $\pi_1N_1$ system, and similarly $H_2 =
H_0(2)
+ H_I(2)$ is the Hamiltonian describing the $\pi_2N_2$ system. The Hamiltonian
describing pion $\pi$ and its interactions with the nucleons can be similarly
written as $H^\pi = H^\pi_0 + H^\pi_I(1) + H^\pi_I(2)$.

Thus, given any Hamiltonian $H=H_0+H_I$ for which we would like to calculate
the
\NN\ OPE potential, we can proceed by replacing $H$ with the sum
$H' = H_1 + H_2 + H^\pi$, where each individual Hamiltonian $H_1$, $H_2$, and
$H^\pi$ has free and interaction parts that are of the same form as $H_0$ and
$H_I$, respectively, but each with its own individual fields replacing the
corresponding ones of $H$.

Although it may be possible to provide a general formulation, it is more
convenient to illustrate the procedure to follow by taking the usual
model where the interaction is given by a three-point \piNN\ vertex.

As we shall not explicitly need to use any details of the model for $H_1$ and
$H_2$, we specify the model in terms of the Hamiltonian involving the pion
$\pi$:
\bea
H_0^\pi &=& \int d\k\, \omega_k\, a^\dagger_\pi(\k) a_\pi(\k) \\
H_I^\pi &=& \int d\k\,  a^\dagger_\pi(\k) J_N(\k)  + H.c. \eqn{Hpi_I}  \\
J_N(\k) &=& \int d\p\,d\p' \delta(\p+\k-\p') \frac{1}{\sqrt{\omega_k}}
F_0(\p,\p') a^\dagger_N(\p) a_N(\p')
\eea
where $N$ can be either $N_1$ or $N_2$, in which case $H_I^\pi$ needs also to
be
labelled accordingly. Note the relations \bea  \left[ a_{\pi}(\k),H_0^\pi
\right]
&=& \omega_k a_{\pi}(\k)  \eqn{aH0_comm} \\ \left[ a_{\pi}(\k),H_I^\pi\right]
&=&  J_N(\k)   .         \eqn{aHI_comm}
\eea
Consider now the perturbation expansion of the full \NNNN\ Green function
with respect to the interactions $H_I^\pi(1)$ and $H_I^\pi(2)$:
\bea \lefteqn{ \la \p'_1,\p'_2 | \frac{1}{E^+-H'} |\p_1,\p_2 \ra  =
\la \p'_1,\p'_2 | \frac{1}{E^+-H_1-H_2-H_0^\pi} H^\pi_I(1) } \nn
&&\frac{1}{E^+-H_1-H_2-H_0^\pi} H^\pi_I(2) \frac{1}{E^+-H_1-H_2-H_0^\pi}
|\p_1,\p_2 \ra  + \ldots
\eea
where the term of order $H_I^\pi(1) H_I^\pi(2)$ has been singled out of the
complete perturbation series, as it is just this term that coincides with the
exact OPE potential specified by the original Hamiltonian $H$. Note how the
introduction of the three Hamiltonians $H_1$, $H_2$, and $H^\pi$, enables us
to treat meson exchange perturbatively, while nucleon dressing is treated
non-perturbatively.

Now consider only this OPE term. Replacing
$H^\pi_I(1)$ and  $H^\pi_I(2)$ by the integral of \eq{Hpi_I}, and then using
\eq{aH0_comm}, we obtain that
\bea \lefteqn{ G_{12}^{OPE} = \int d\k \, \la
\p'_1,\p'_2 | \frac{1}{E^+-H_1-H_2} J^\dagger_{N_1}(\k)
\frac{1}{E^+-H_1-H_2-\omega_k} J_{N_2}(\k) } \hspace{8cm} \nn
&&\frac{1}{E^+-H_1-H_2} |\p_1,\p_2 \ra           \hspace{2cm}
\eqn{G_OPE_midway}
\eea where we have used that $H^\pi_0$ acting on two-nucleon states gives zero,
and where one $\k'$-integral has been eliminated using $[a_\pi(\k'),
a^\dagger_\pi(\k)] = \delta(\k'-\k)$. The essential step comes at this stage
when we recognize that $[H_1,H_2] = [H_1,J_{N_2}]=[H_2,J_{N_1}]=0$ which
enables
us to write \eq{G_OPE_midway} in terms of two contour integrals:
\bea
\lefteqn{ G_{12}^{OPE} = \left(-\frac{1}{2\pi i}\right)^2 \int d\k\,dz\,dz' \la
\p'_1| \frac{1}{z'^+-H_1} J^\dagger_{N_1}(\k) \frac{1}{z^+-H_1} |\p_1 \ra
\frac{1}{z'-z-\omega_k} } \hspace{5cm} \nn
&& \la \p'_2 | \frac{1}{E^+-z'-H_2}J_{N_2}(\k)
\frac{1}{E^+-z-H_2} |\p_2 \ra       \hspace{2cm}     \eqn{G_OPE_soon}
\eea
where the matrix element factors into two, one factor for each nucleon. Because
of this factorization, and because all Hamiltonians have the same form as $H$,
we may now drop the nucleon labels in \eq{G_OPE_soon}.

The matrix elements in \eq{G_OPE_soon} define the two-energy vertices
\be
\delta(\p'+\k-\p) f(\p',\p,z',z) = g^{-1}(z',\p') \, \la \p'|
\frac{1}{z'^+-H} J_{N}(\k) \frac{1}{z^+-H} |\p \ra \, g^{-1}(z,\p).
\ee
Substituting this definition into \eq{G_OPE_soon} results in the expression
of \eq{tilde_V_OPE}.

Using \eqs{aH0_comm} and (\ref{aHI_comm}), it is straightforward to
show that
\bea
 \lefteqn{ \la \p'| \frac{1}{z'^+-H} J_{N}(\k) \frac{1}{z^+-H} |\p \ra \ } \nn
&=&
\la \p'\k| \frac{1}{z^+-H} |\p \ra \
+ (z-z'-\omega_k) \la \p'| \frac{1}{z'^+-H} a_{\pi}(\k) \frac{1}{z^+-H} |\p
\ra \   .    \eqn{one_two}
\eea
Recognizing that the first term on the rhs is the usual one-energy vertex
function, \eq{one_two} shows that the one- and two-energy vertex functions
coincide for on-mass-shell pions.
\bigskip\medskip

\centerline{\bf SUMMARY}
\bigskip

Attempts to formulate few-body equations for the \piNN\ system have a long
history. Yet all attempts from the very first bound state model of the 1960's
to the sophisticated \NNpiNN\ models of the 1990's
had one basic feature in common, each attempted to simplify the field
theory in question by truncating to some maximum number of pions.
Unfortunately,
this truncation results in an inconsistent treatment of nucleon dressing
with serious consequences for practical calculations. The \NNpiNN\
equations, for example, suffer from a renormalization problem which effectively
reduces the strength of the input \piNN\ vertices, which in turn contributes
to the underestimation of cross sections.

In order to resolve these renormalization problems, we have presented a
completely different approach where the Hilbert space is not truncated to some
maximum number of pions. Instead, the guiding principle has been to work only
with fully dressed vertices and propagators. What this means in practice, is
that all possible disconnected diagrams of quantum field theory have been
retained in our model. To achieve this, we have used convolution integrals to
sum
over all relative time orderings of disconnected graphs. As convolution
integrals have often been used in nuclear physics, especially in the
four-nucleon problem \cite{Hab}, it may appear surprising why the idea has not
been applied sooner to the \piNN problem. This, however, can be understood when
it is realized that the convolutions that we have used are applied to field
theory Hamiltonians, and result in convolution expressions for {\em
energy-dependent} "potentials". This is quite different from the more usual
convolutions of nuclear physics which apply {\em only} for {\em
energy-independent} potentials.

The convolution \piNN\ equations are unitary and do not suffer from
renormalization problems. These are the two crucial attributes that are
necessary for a theoretically consistent description. Indeed, we should
emphasize that the downfall of the \NNpiNN\ equations is {\em not} because the
two-nucleon propagator is underdressed, rather, it is simply because the
equations themselves do not have both the properties of unitarity and correct
renormalization.

The only approximation made in deriving the convolution \piNN\ equations is
the neglect of connected three-body forces. This is, of course, also an
assumption in essentially every model in nuclear physics. Drawing on the bulk
experience of the field, such three-body forces are very likely to be small.
However, to obtain a more quantitative assessment, explicit calculations
of connected three-body force contributions are needed. With this goal in mind,
we have extended the convolution idea also to connected diagrams.

Two methods have been presented to derive convolution integrals that sum all
the time-orderings of a connected graph. The first method, involving the
two-time Green function of relativistic field theory, is useful to obtain
the form of the convolution integral in a quick and straightforward manner.
However, this method works only in the case when dressed and bare vaccua are
equal. The second method, is an extension of the one used by us to derive
the convolution formula for disconnected diagrams \cite{KB_PRC}, and involves
replacing the true Hamiltonian $H$ by a sum of Hamiltonians, each having the
same form as $H$, but applying only to singled out particles. This method
works without any assumptions and is therefore more general than the
one involving the two-time Green function.

With the convolution formula for connected diagrams derived, we have calculated
the fully dressed \NN\ OPE potential, since part of this potential includes
connected three-body forces. For this dressed OPE potential, we find that the
corresponding connected three-body force is negligible. Thus we can conclude,
that at this stage, the convolution \piNN\ model is a candidate for effectively
summing the whole of time-ordered perturbation theory for the \piNN\ system.

\end{document}